\documentclass{vgtc}                          
\let\ifpdf\relax

\usepackage{mathptmx}
\usepackage{graphicx}
\usepackage{times}
\usepackage{subcaption}
\usepackage{booktabs}

\onlineid{0}

\vgtccategory{Research}

\vgtcinsertpkg

\title{Effects of Realism and Representation on Self-Embodied Avatars in Immersive Virtual Environments}

\author{Rafael Kuffner dos Anjos, Jo\~ao Madeiras Pereira}

\abstract{

Virtual Reality(VR) has recently gained traction with many new and ever more affordable devices being released.
The increase in popularity of this paradigm of interaction has given birth to new 
applications and has attracted casual consumers to 
experience 
VR.
Providing a self-embodied representation (avatar) of users' full bodies inside 
shared virtual spaces can improve the VR experience and 
make it more engaging to both new and experienced users . 
This is especially important 
in fully immersive systems, where the equipment completely occludes the real world 
making self awareness problematic.
Indeed, the feeling of presence of the user 
is highly influenced by their virtual representations, even though small flaws could lead to uncanny valley side-effects. 
Following previous research, we would like to assess whether 
using a third-person perspective could also benefit the VR experience, 
via an improved spatial awareness of the user's virtual surroundings. 
In this paper we 
investigate 
realism and perspective of self-embodied representation in VR setups in natural tasks, such as walking and avoiding obstacles.
We compare both First and Third-Person perspectives with three different levels of realism in avatar representation. These range from a stylized abstract avatar, to a ``realistic" mesh-based humanoid representation and a point-cloud rendering.
The latter 
uses data captured via depth-sensors and mapped into a virtual self inside the Virtual Environment.
We present a throughout evaluation and comparison of these different representations, describing a series of guidelines for self-embodied VR applications.
The effects of the uncanny valley are also discussed in the context of navigation and reflex-based tasks.
} 

\CCScatlist{ 
  \CCScat{K.6.1}{Management of Computing and Information Systems}%
{Project and People Management}{Life Cycle};
  \CCScat{K.7.m}{The Computing Profession}{Miscellaneous}{Ethics}
}

\begin{document}


\firstsection{Introduction}

\maketitle


The use of Virtual Reality (VR) systems can improve efficiency on many tasks in different areas of expertise such as engineering, medicine and architecture.
But a common problem in these setups is the need to use optical tracking systems to locate objects in the 3d shared space.
This is even more problematic to track the full-body of a user, needing them to wear reflexive rigid body markers for each of its joints for correct representation inside the environment.
The use of Kinect Sensors overcame this problem by providing non-intrusive full-body skeleton tracking and can also extract their image and map it to its virtual self.

When using a Head-Mounted Display system, the real-self of the user is completely occluded, which diminishes the sense of presence in this shared virtual space.
The use of a self-embodied representation (avatar) is one approach to overcome this issue~\cite{slater1994body}.
The use of an avatar gives a reference of recognizable size and a connectedness to the virtual environment  \cite{interrante2006distance,ries2008effect, ribeiro2016moco}, even though may still cause distance underestimation ~\cite{renner2013perception}. 
An important aspect of measuring the effectiveness of a VR experience is the sense of embodiment.
This concept is constitutive of the sense of presence in VR and affects the way one interacts with virtual elements~\cite{kilteni2012sense}. 
The sense of embodiment of an avatar is subdivided in three components: the sense of agency, i.e.  feeling of motor control over the virtual body; (ii) the sense of body ownership, i.e.  feeling that the virtual body is one’s own body; and (iii) self-location, i.e. the experienced location of the self.

Some aspects are known to influence the sense of embodiment of the avatar, namely the realism of the representation and the perspective which is viewed.
Regarding realism in avatars, a known problem is the uncanny valley ~\cite{mori2012uncanny}, which also affects embodied representations ~\cite{lugrin2015avatar,lugrin2015influence}.
Changing the perspective which the avatar is viewed could also positively contribute to the VR experience~\cite{debarba2015characterizing,ehrsson2007experimental} without compromising user's sense of embodiment. 
Normally, the user is viewed in its own perspective (known as First-Person Perspective).
Another possibility is the use of Third-Person perspective (3PP) where the virtual camera is positioned behind the user, allowing them to view its own full-virtual body.
This type of representation is widely used in games for improving spatial awareness in conventional displays.

In this paper we further study the influence of perspective (1PP and 3PP) and realism of the representation of self-embodied avatars in the sense of embodiment of users in Virtual Reality setups. 
For that, we use three different representations following the known Uncanny Valley effect~\cite{mori2012uncanny} varying the level of realism of each representation, from an abstract to a realistic humanoid representation.
The first is an abstract representation that uses spheres and cylinders to represent parts of the body.
The second is a realistic mesh avatar that is deformed according to the tracking information.
The third representation is a low cost point-cloud based avatar, which extracts information from one's real-self and maps it into the virtual environment. 
For comparing sense of embodiment, efficiency and easiness of use of each of the representations we designed and evaluated four different natural tasks based on  previous work~\cite{salamin2010quantifying}.
The First three tasks were navigation tasks, where users needed to walk while avoiding obstacles in the Virtual Environment. In the latter, users needed to use their reflexes to catch objects thrown in in their direction.
In the following sections we present the related work on user-representation, describe the test measures, the results obtained and propose a set of guidelines for Self-Embodied Virtual Reality applications.




\section{Related Work}

In this section we present and discuss the related work regarding user representation. First, we introduce and define virtual avatars and concepts associated.
Following, a brief review of point cloud visualization techniques is discussed, and their use and viability in Virtual Reality setups

\subsection{Virtual avatars}


An important part of the experience is how users are represented on the virtual scene. 
As opposed to CAVE-like systems, the use of Head-Mounted Display technology occludes users' real self, compromising the overall virtual-reality session. 
A way of overcoming this problem is by using a fully-embodied representation of the user within the virtual environment ~\cite{slater2010simulating}.
A virtual body can supply them with a reference of recognizable size and a connectedness to the virtual environment  \cite{interrante2006distance,ries2008effect}, even though studies indicate that the use of a virtual body setups may still cause distance underestimation ~\cite{renner2013perception}. 
The feeling of presence is related to the concept of proprioception, which is the ability to sense stimuli arising within the body regarding position, motion, and equilibrium. 
The sense of embodiment into an avatar is constitutive of the sense of presence in virtual reality (VR) and affects the way one interacts with virtual elements~\cite{kilteni2012sense}. 
This concept is subdivided in three components: the sense of agency, i.e.  feeling of motor control over the virtual body; (ii) the sense of body ownership, i.e.  feeling that the virtual body is one’s own body; and (iii) self-location, i.e. the experienced location of the self.

The level of realism of the avatar also plays an important part on the VR experience and how it relates to the sense of embodiment of an user. 
A common problem on this matter is the uncanny valley \cite{mori2012uncanny}, which states that the acceptability of an artificial character will not increase linearly in relation to its likeness to human form. Instead, after an initial rise in acceptability there will be a pronounced decrease when the character is similar, but not identical to human form. Additionally, Piwek et al.\cite{piwek2014empirical} state that the effect of realism in the deepest part of the valley become more acceptable when it is animated. 
Works by Lugrin et al. \cite{lugrin2015avatar,lugrin2015influence} also state that the uncanny valley also affect the feeling of presence and embodiment of avatars in first person perspective (1PP) when viewed through a head-mounted display.

Another possibility when using a self-embodied avatar of the user is changing the perspective which the avatar is viewed. 
This approach is normally used on video-games to increase user's spatial awareness when navigating and interacting on the scene. 
The sense of body ownership is also possible when using artificial bodies (in real scenarios) and avatars (in virtual environment scenarios) in immersive setups. 
A classical extra-corporeal experience is known by Rubber Hand Illusion (RHI) ~\cite{botvinick1998rubber}. 
In this illusion, a subject is made to believe a rubber hand is in fact his own hand, which is hidden from view, to the point of pulling his own hand away if the rubber hand is attacked.
This illusion has similar effects in Virtual Reality setups, which is called Virtual Hand illusion, and can be induced by visuotactile ~\cite{slater2008towards} and visuomotor synchrony ~\cite{sanchez2010virtual,yuan2010rubber}. 

The Rubber-hand Illusion has also proven to work with full-body embodiment. 
Ehrrson et al. ~\cite{ehrsson2007experimental} proves this by streaming a the image of the body of the participant with an image of his body in a third-person perspective using an stereoscopic camera.
Leggenhager et al. ~\cite{lenggenhager2009spatial} confirms this by using a similar setup to prove that when using a Third-person perspective behind users bodies, users felt to be located there they saw the virtual body to be.
In VR, the usage of orthogonal third person viewpoints has been explored and was for instance recommended to help setting the posture of a motion controlled virtual body ~\cite{boulic2009scaling}. 

Further work by Salamin et al. use an augmented-reality setup with a displaced camera and an HMD to show that the best perspective depends on the performed action: first-person perspective (1PP) for more precise object manipulations and third-person perspective (3PP) for moving actions.
Work by the same work also shown that the users preferred the use of the 3PP in comparison with 1PP and needed less training in a ball catching scenario~\cite{salamin2010quantifying}. Further work by Kosch et al. ~\cite{kosch2016exploring}, find that the preferred viewpoint in a 3PP is behind user's head, providing a real life third person experience.
Distance underestimation is also present when the avatar is seen on a third person perspective ~\cite{salamin2010quantifying}.

\subsection{Point cloud visualization}

The main challenge when dealing with point-cloud visualization is the unstructured nature of the data and its sparsity.
Rendering point clouds with point primitives has several drawbacks when compared to other techniques (e.g., background/foreground confusion, loss of definition on close-ups), when confronting a low resolution scenario \cite{1175093}.
Katz et. al \cite{Katz:2007:DVP:1276377.1276407} solved the problem of background foreground confusion by estimating the direct visibility of sets of points.
However, in a mixed visualization scenario such as the one applied on this work, confusion still exists between the rendered points and for the body and the mesh-based environment. 

Surface reconstruction is the standard approach to visualize point cloud data  \cite{fabio2003point} with several successful techniques estimating the original surfaces from point sets \cite{gopi2000surface,Kazhdan:2006:PSR:1281957.1281965}. 
A single depth stream can be easily remeshed using delaunay triangulation \cite{CGF:CGF439}, or multi-stream fusion can be performed such as shown in the work of Dou et al. \cite{Dou:2016:FRP:2897824.2925969}.
While single stream triangulation can be performed in real time, multi stream fusion can be a very consuming task that requires specialized hardware for processing data.   

Screen-aligned splats \cite{westover1991splatting} have been proposed as a more efficient alternative to polygonal mesh rendering \cite{Rusinkiewicz:2000:QMP:344779.344940}, and are easily implemented in an interactive system, being the to go approach for visualization of real time data. 
Authors claim to have a comparable visual appearance to closed surfaces for visualization uses \cite{Botsch:2002:EHQ:581896.581904}. 
Although surface aligned splats \cite{preiner2012auto,SPBG:SPBG05:017-024,Zwicker04perspectiveaccurate} which create a better approximation of the surface, normal estimation in real-time can be also a costly operation. 
Splatting has been used in the past for point cloud visualization of environments \cite{7823453}, but not in a real time reconstruction of the users body.

\section{Test Design}


We further study the effects of realism and perspective in natural tasks.
In this section we describe the main aspects of designing the test experience regarding user representation and the design of the task. 
The following subsections present the task concept, the avatar representations used and the setup used on the test task.

\subsection{Setup}

\begin{figure}[tbh]
    \captionsetup[subfigure]{position=b}
    \centering
    \subcaptionbox{Reality. Kinect sensors marked with yellow, HMD with blue. \label{fig:real}}{\includegraphics[width=\columnwidth]{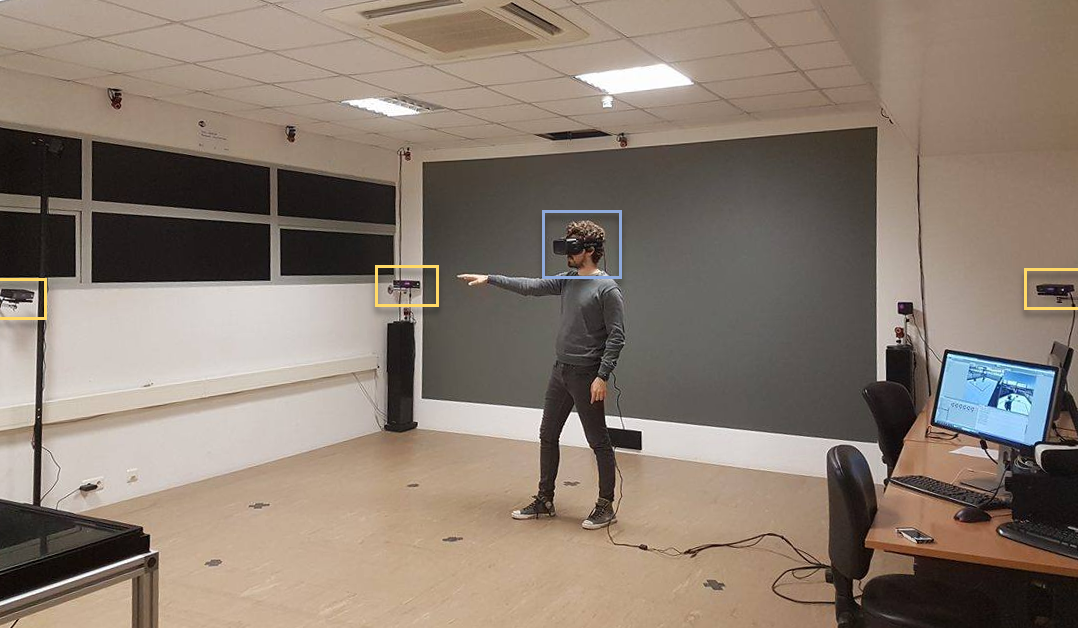}}
    \subcaptionbox{Virtual World with point cloud avatar \label{fig:virtual}}{\includegraphics[width=\columnwidth]{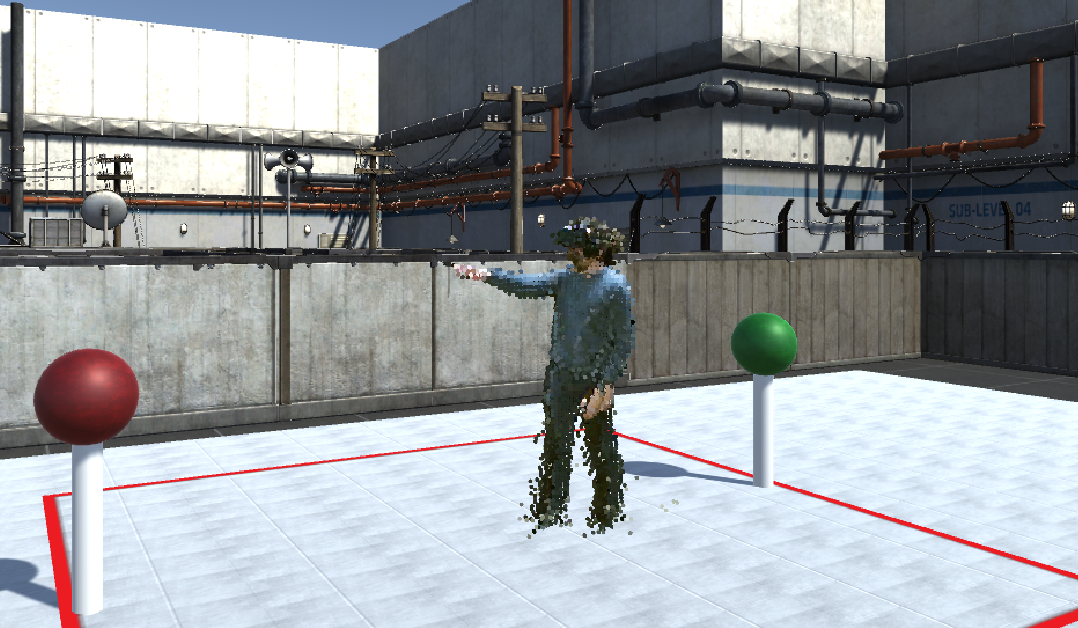}}
    \caption{The setup used for our study. Figure \ref{fig:real} shows the laboratory and one user performing a test, and Figure \ref{fig:virtual} the mapping to the virtual world.}
    \label{fig:setup}
\end{figure}

A wide-baseline setup was used due to two main reasons; firstly the fact that the kinect sensor has a limit on its effective range (0.4m to 4.5m, with skeletons losing reliability starting on 2.5m), and in order to properly evaluate a navigation task, a bigger space was needed.
When the user is at the limits of the sensors operating range, the quality of the experience would be compromised, so a wide-baseline setup guarantees the whole body of the user is always visible by at least one camera.
Secondly, since a third person perspective is presented as one interaction paradigm, the whole body of the participant must be visible at all times in order to avoid holes in the representation.
A narrow baseline or single sensor setup would capture just half of the participant's body, greatly compromising the experience. 

The five sensors are fixated on the walls of the laboratory where the study was being held, covering an area of approximately 4 x 4 meters. 
Since the proposed navigation tasks were mainly performed on a line between the two goals, we mapped the environment in such a way so during the execution of the tasks, the participant was always facing a sensor so his hands were always visible on the first person perspective, and back on the third person perspective. The physical setup chosen for our study can be seen on figure \ref{fig:setup}. 

\subsection{User Representations}
Regarding user representation, we chose three different user representations, following the known Uncanny Valley effect.
Camera positioning in third-person perspective is based on previous work by Kosch et al.~\cite{kosch2016exploring}, where the camera is positioned above user's head for improved spatial awareness.

In all the used representations, the Kinect's joints position and rotation are mapped directly into the avatars using direct Kinematics.

\subsubsection{Abstract}
The first avatar is a simplified avatar representation which is composed by abstract components.
Spheres were used for each joint and the head, and cylinders for each bone connecting joints.
Only the joints that are tracked by the Microsoft Kinect were represented.
Figures \ref{fig:repa} and \ref{fig:repb} how this representation on the First and Third Person Perspectives (1PP and 3PP), respectively.

\subsubsection{Mesh}
The second representation is a realistic mesh avatar resembling a human being. 
This representation did not include animation for individual fingers, since they are not tracked by the Kinect sensor. 
Figures \ref{fig:repc} and \ref{fig:repd} show this representation on the First and Third Person Perspectives (1PP and 3PP), respectively.

\subsubsection{Point Cloud}

This body representation is based on a combination of separate streams of point clouds from Microsoft Kinect sensors.
Each individual sensor first captures the skeletal information for each human in its field of view. 
Following, a point cloud is created from the combination of depth and color values seen by the camera, and points relevant to the users are segmented from the background.

At several points of the interaction, different cameras will be sending very similar information, and due to the timely constrained nature of our problem, integration of different streams or redundancy resolution is not performed. 
We chose to implement a simple decimation technique that takes into account the what body parts are more relevant to the task at hand. 
By using the captured skeleton information, we attribute different priorities to each joint according to user defined parameters.
For the Virtual reality scenario, information about hands was found to be more valuable, and head information was discarded. 

For each point $p_i$ in the segmented cloud, we calculate its euclidean distance $d = |p_i-j_k|$ to the joints $j_k$ marked as relevant. 
If $d$ is smaller to a threshold value $t$ (set according to the user's body size.), this point is marked accordingly as a high quality point for transmission. 
Points not marked as high quality, are sampled on a lower frequency (according to application parameters) in order to reduce redundancy of data on areas less relevant to the task.

Since each sensor can be associated with a single computer, we transmit both the skeleton and point cloud data through the network to the host PC where the application is running.
For each point, we transmit $(x,y,z,r,g,b,q)$ which are the points 3D coordinates, color, and one bit indicating high or low quality. 
This last bit is necessary in order to adjust the rendering parameters in the host application.
While higher quality points are more tightly grouped, requiring smaller splat sizes, under-sampled regions must use bigger splats to create closed surfaces. 

\begin{figure}[t!]
    \captionsetup[subfigure]{position=b}
    \centering
    \subcaptionbox{1PP Abstract Avatar  \label{fig:repa}}{\includegraphics[width=.45\columnwidth]{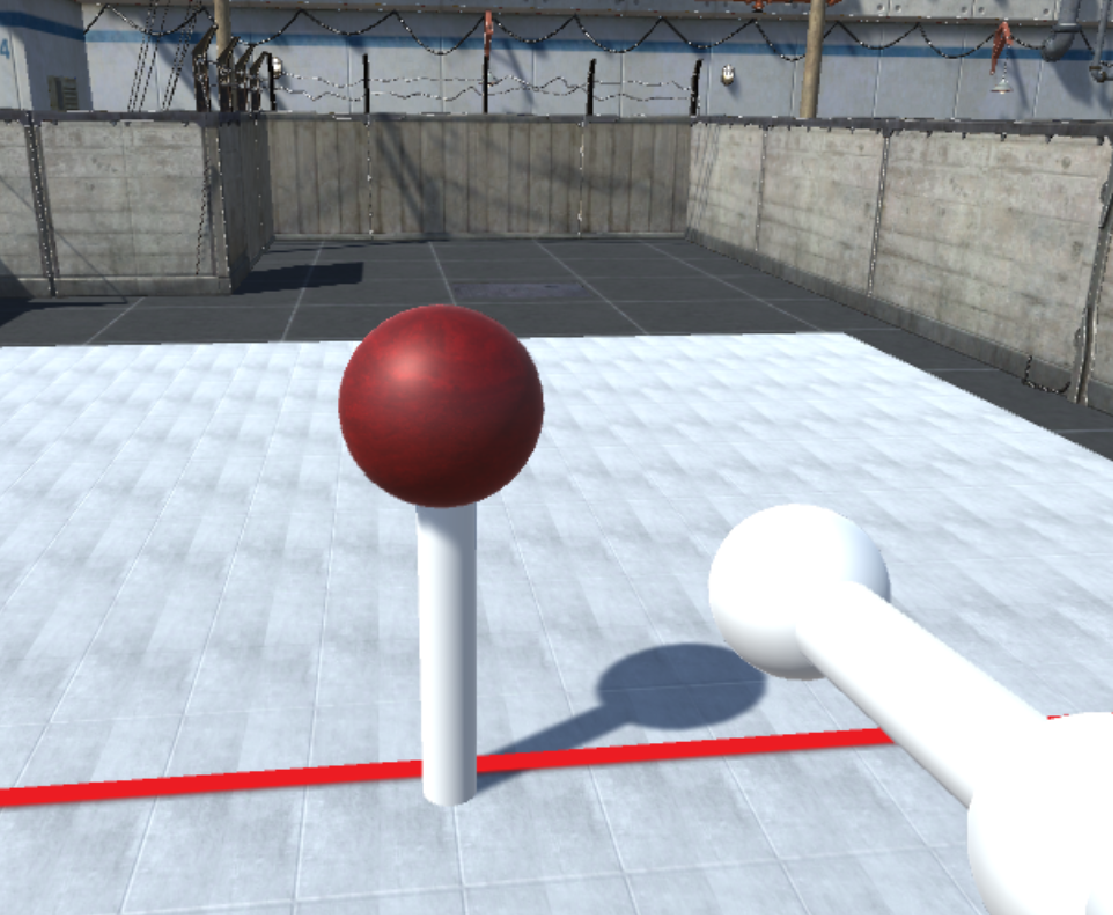}}
    \subcaptionbox{3PP Abstract Avatar  \label{fig:repb}}{\includegraphics[width=.45\columnwidth]{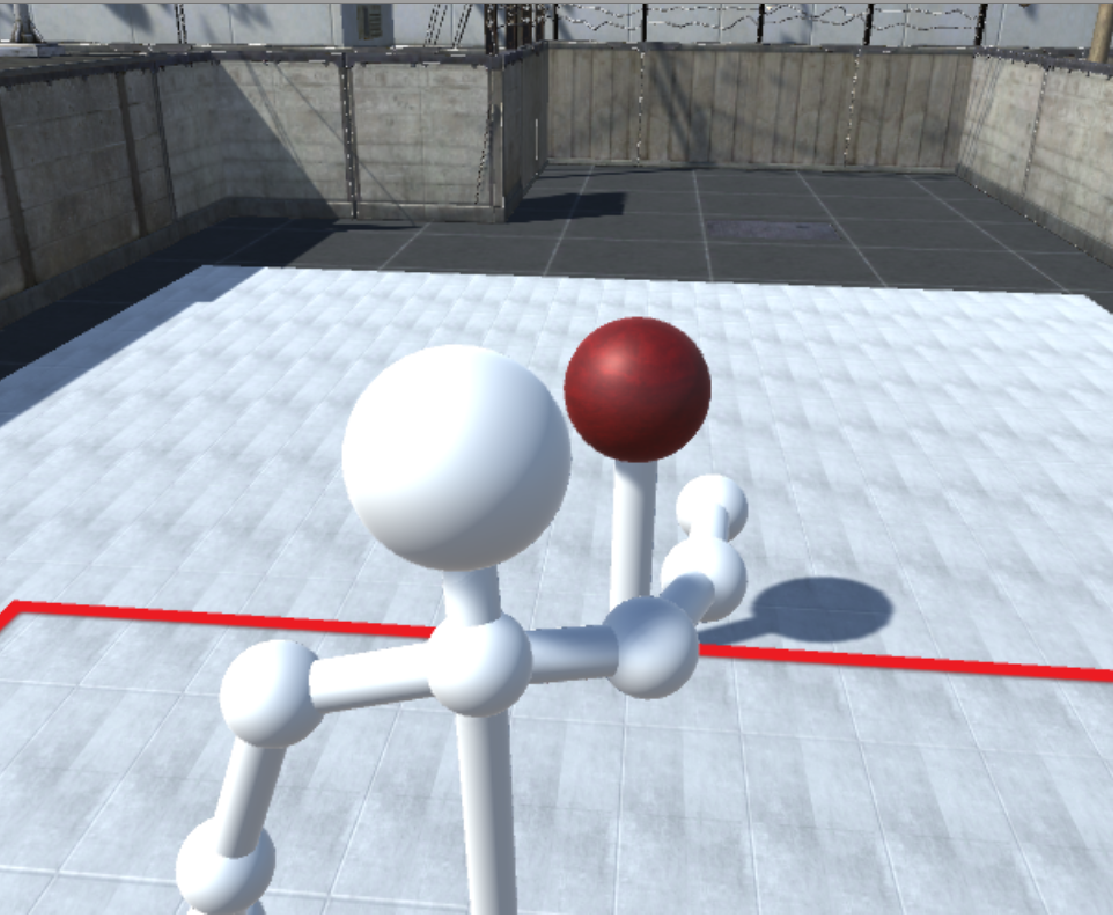}}
    \subcaptionbox{1PP Mesh Avatar \label{fig:repc}}{\includegraphics[width=.45\columnwidth]{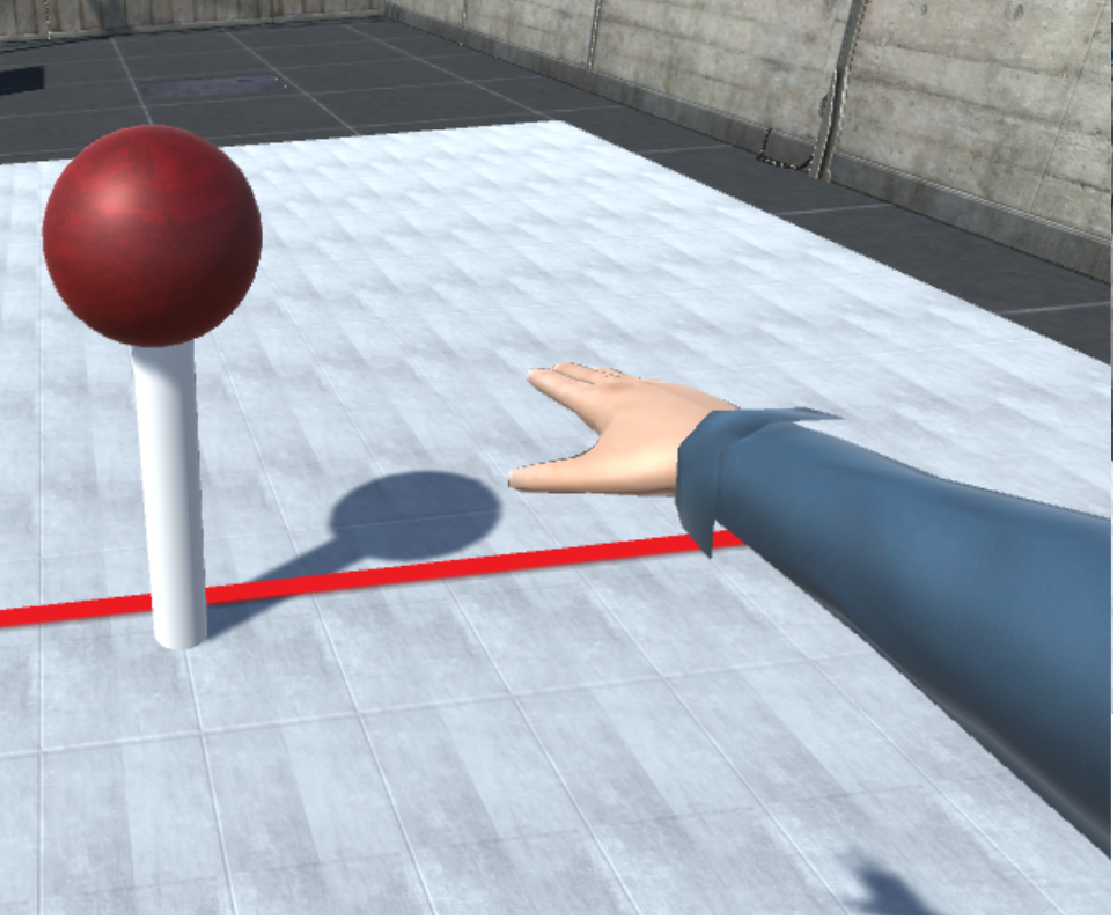}}
    \subcaptionbox{3PP Mesh Avatar \label{fig:repd}}{\includegraphics[width=.45\columnwidth]{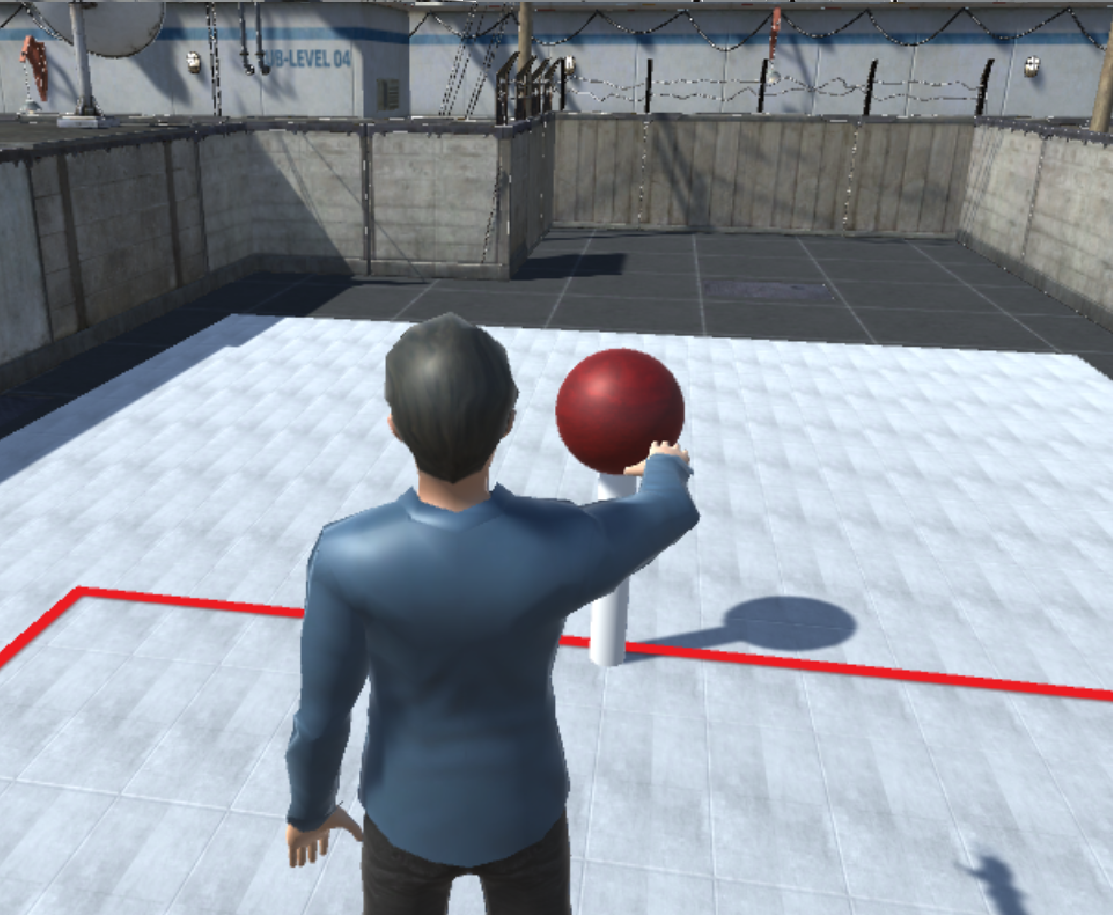}}
    \subcaptionbox{1PP Point-Cloud Avatar \label{fig:repe}}{\includegraphics[width=.45\columnwidth]{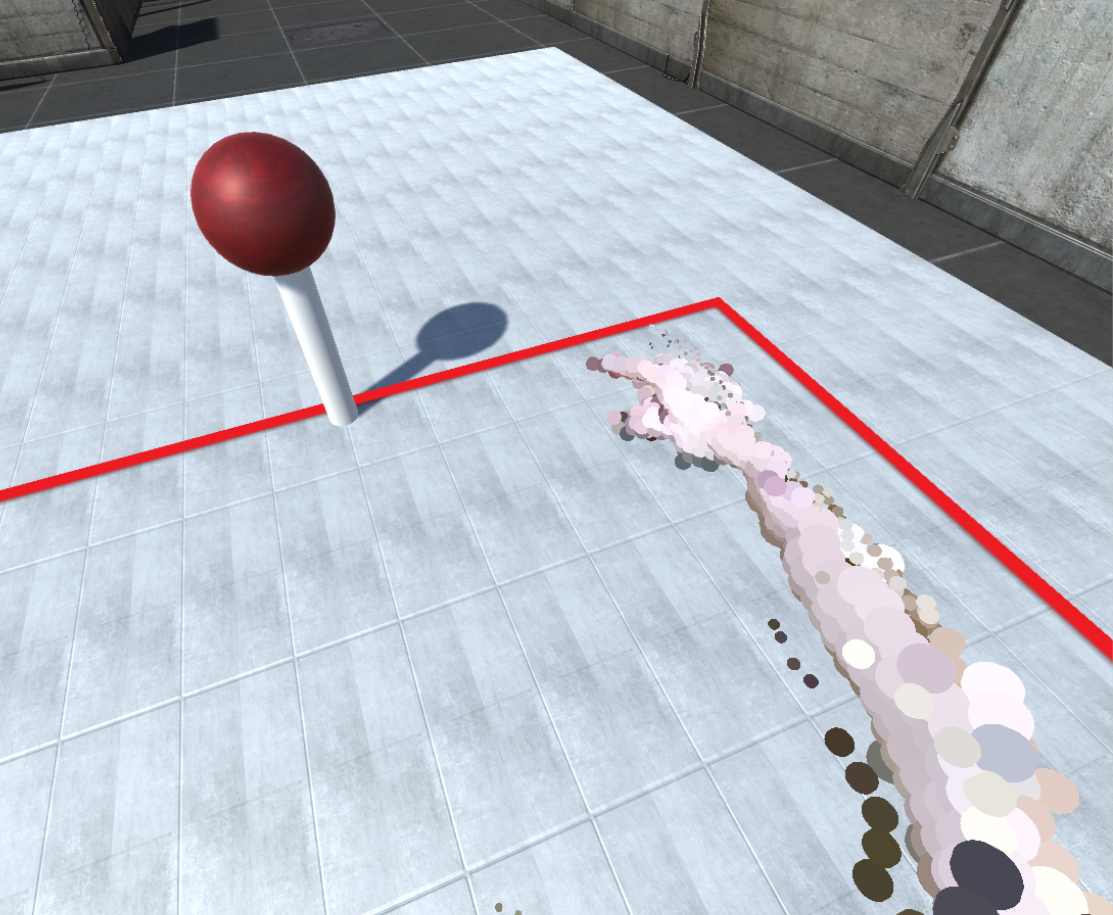} }
    \subcaptionbox{3PP Point-Cloud Avatar \label{fig:repf}}{\includegraphics[width=.45\columnwidth]{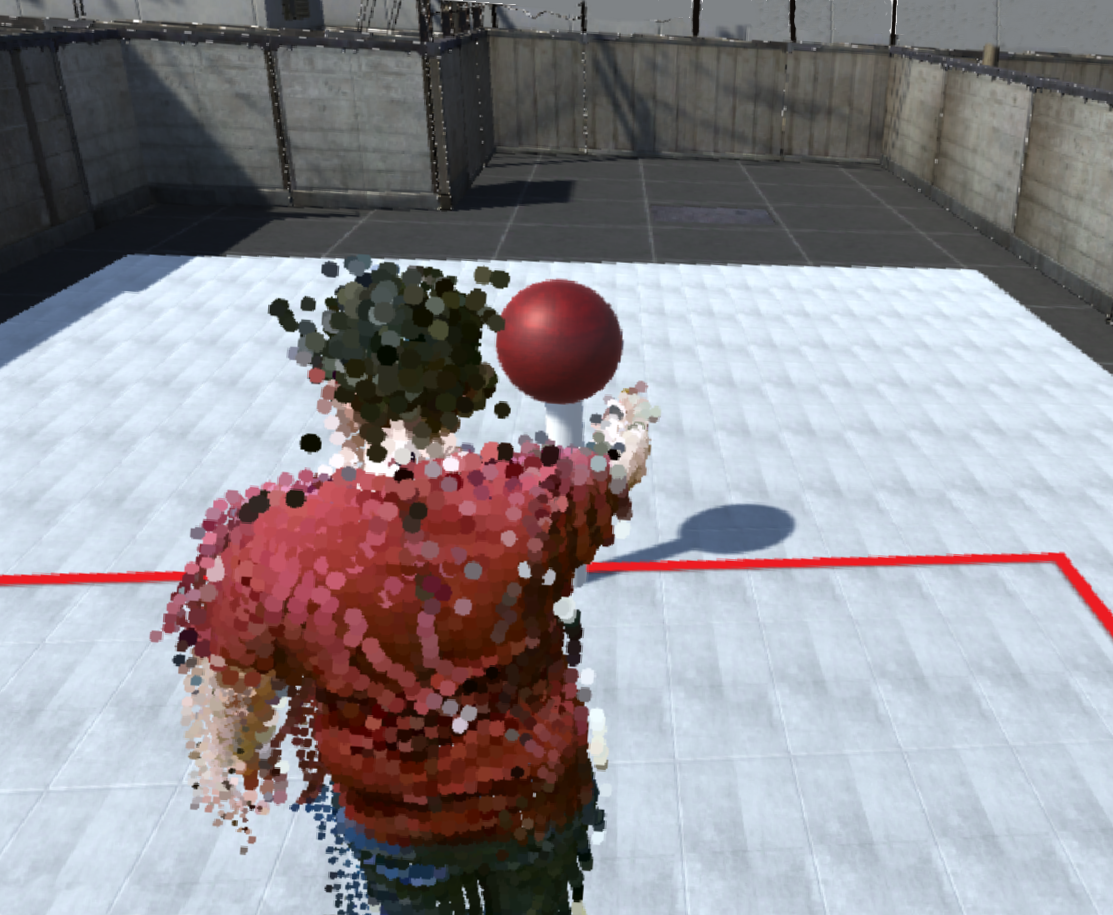}}
    \caption{Self-representations used on our study.}
    \label{fig:representations}
\end{figure}

Each sensor position is previously configured on the host computer after a calibration step.
Data read through the network is then parsed and rendered in the environment using surface-aligned splats.
For interaction purposes, the transmitted skeleton information is used.
We are able to successfully parse and render the transmitted avatars at 30 frames per second, allowing for a clean interaction on the user side. Figures \ref{fig:repe} and \ref{fig:repf} show this representation on the first and third person points of view.

\subsection{Methodology}

The test was divided into eight stages: 1) introduction to the study and application of pre-test questionnaire; 2) explanation about the tasks and each of the users representations 3) adjustment of the device for comfort; 4) calibration procedure; 6) task execution; 7) application of post-test questionnaire; 8) and a semi-structured interview.

At first, we explained the test objectives. Then, the users completed a pre-use questionnaire to raise the participants profile regarding previous experience with related technologies (HMDs, virtual avatars, etc). 

Subsequently, we showed a brief description of the tasks and representations used. 
Furthermore, we presented the users with the calibration task. This procedure was performed to calibrate the tracking system between the HMDs and the depth-sensors.
Then, in order to familiarize the user with the procedures, users performed a task in a training scenario, where they could freely explore the virtual environment  and familiarize themselves with the setup and each of the representations.   

After performing the training task, the users reached a fixed object on the environment and the users performed the test task. Then a questionnaire was given to the users for some user experience issues. 
These steps were done for each of the combination of the test conditions(perspective and representation) of a total of 12 permutations. The order of the avatar representation was changed in every test, following a Latin square arrangement,  to avoid biased results.

\subsection{Virtual Environment}

The selected environment  is based on the  Stealth Scene, which was obtained on the Unity Asset Store\footnote{http://unity3d.com/store}.
This scene was modified to remove visual clutter, for not interfering with the goals of the test, stealing user's attention.

We also included in the environment a representation of the Kinect's tracking limits with a red square, where the user could walk freely. 

\subsection{Tasks Description}

Because of the reduced size of the tracked space (4 meters by 4 meters), we decided to divide the test in four tasks.
For each of the tasks, the user need to reach a specific point at the end of the tracked space, where he needs to reach a colored sphere (either red or green) with their hand to pass to the next Task. 
The Tasks were chosen based on natural Tasks such as walking,avoiding obstacles and catching moving objects based on previous work~\cite{salamin2010quantifying}.

After that, he turns his body to face the next task, until the test ends. 
In the following subsections we present and explain in further detail each of the Tasks.

\subsubsection{First Task}

In this task, after reaching the green object the user turns his body and needs to go around the barrels, first by the right, then by the left until they reach the read target object. Figure~\ref{fig:firstTask} illustrates the first Task.

\subsubsection{Second Task}

In the second task, the user needs to avoid each of the yellow bars by raising their feet (or jumping) until they reach the green target Object (Figure~\ref{fig:secondTask}).

\subsubsection{Third task}

The user needs to avoid the yellow bar by going under it.
This bar is adjusted according to the user's height, which is estimated using the distance between the head and the toe when the user starts the test. 
The bar is placed 12 centimeters less than the total height of the user (Figure~\ref{fig:thirdTask}).

\subsubsection{Fourth task}

\begin{figure}
\centering

\begin{subfigure}[t]{0.45\columnwidth}
\centering
\includegraphics[width=\columnwidth]{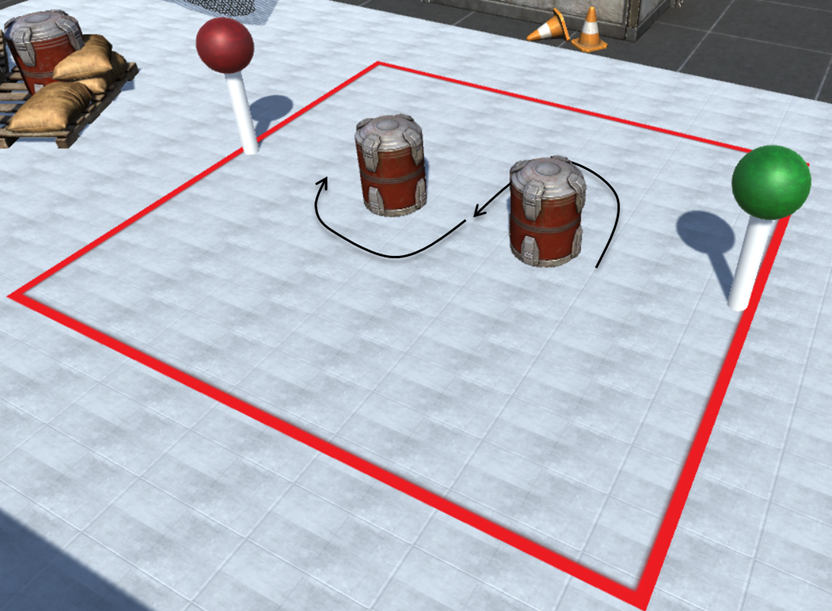}
\caption{First Task}
\label{fig:firstTask}
\end{subfigure}%
\hfill
\begin{subfigure}[t]{0.45\columnwidth}
\centering
\includegraphics[width=\columnwidth]{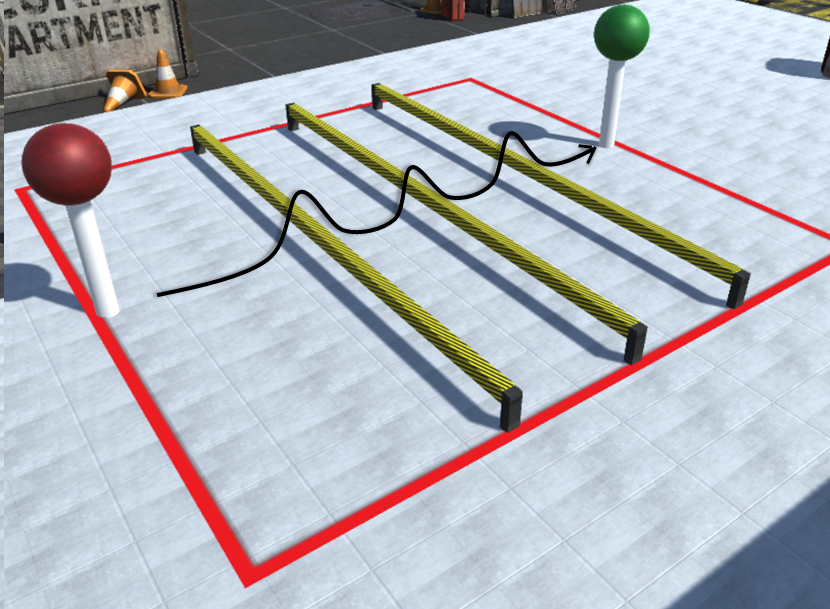}
\caption{Second Task}
\label{fig:secondTask}
\end{subfigure}

\bigskip 

\begin{subfigure}[t]{0.45\columnwidth}
\centering
\includegraphics[width=\columnwidth]{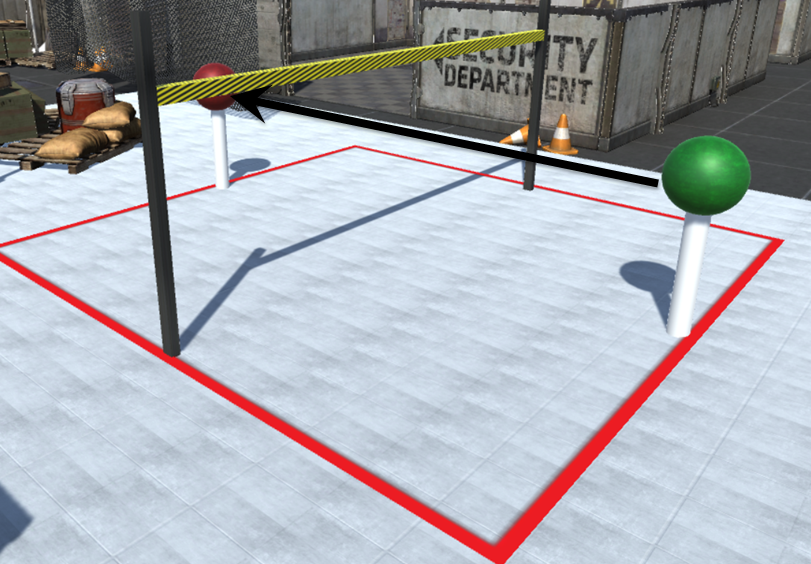}
\caption{Third Task}
\label{fig:thirdTask}
\end{subfigure}%
\hfill
\begin{subfigure}[t]{0.45\columnwidth}
\centering
\includegraphics[width=\columnwidth]{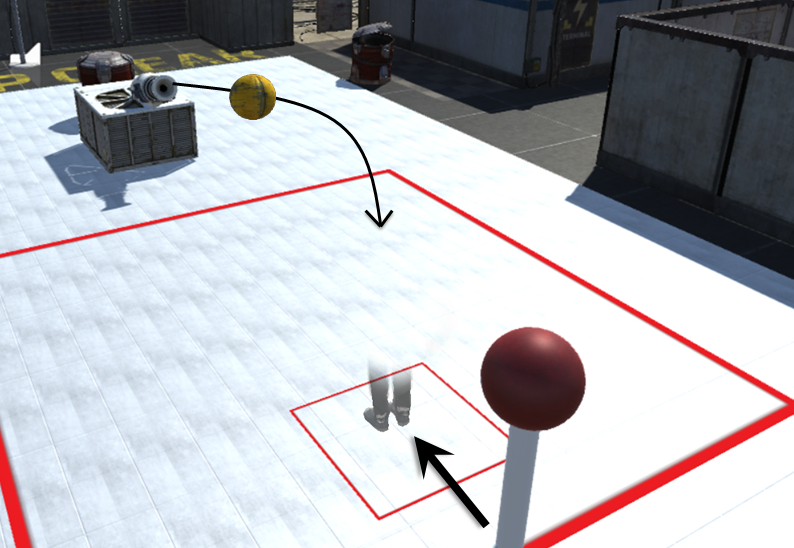}
\caption{Fourth Task}
\label{fig:fourthTask}
\end{subfigure}

\caption{The proposed tasks for our evaluation.}
\label{fig:tasks}
\end{figure}

In the final task the user needs to reach a small red square located in front of the red target object. This task is a Reflex-based task which consists in four balls been thrown towards the user by a cannon (Figure~\ref{fig:fourthTask}).
To succeed in this task they need to stop them using their body. 
All balls are thrown with the same speed, varying only the point they are thrown at.
The first ball is thrown toward the user's chest, second and third towards user's right and left arm.
The fourth ball is thrown at the user's right side but he needs to walk to the right to reach it.

\subsection{Questionnaires}

To gather users profiles, preferences a pre-test questionnaire was conducted. 
For assessing human factors such as comfort, sense of embodiment and satisfaction, a post-test questionnaire was conducted. 
The post-test questionnaire was comprised of a list of 11 statements followed by a Likert Scale of 6 values, where 1 means that the user does not agree completely with statement and 6 means he fully agrees with it, as summarized in Table \ref{tab:questionnaires}.
The first four questions, based on previous work~\cite{debarba2015characterizing}, were used to estimate user's sense of embodiment and each of its sub-concepts : sense of embodiment (Q1), sense of body ownership (Q2) ,sense of self-location (Q3) and also body control (Q4). 
The following questions were made to assess the easiness of each of the subtasks and fatigue. 

In addition to the questionnaire, we conducted a semi-structured interview in order to capture the participants' perceptions about the accomplished tasks, clarify about their answers on the post-test questionnaire and get improvement suggestions.%

\subsection{Participants}

For this test we chose 24 participants, 5 of which were female. The ages of the users varied from 21 to 35 years.
Regarding experience, the majority of the users had previous experience with 3d applications such as games and modelling systems.
The majority of them with experience with previous experience with Head-Mounted Displays (18 participants, or 75\%) and 21 with previous experience with Kinect usage.



\begin{table*}[t]
\centering
\caption{Summary of the questionnaires. * indicates statistical significance}
\label{tab:questionnaires}
\begin{tabular}{@{}lllllll@{}}
\toprule
 & \multicolumn{3}{l}{First-Person Perspective} & \multicolumn{3}{l}{Third-Person Perspective} \\ \midrule
& Abstract & Mesh & Point & Abstract & Mesh & Point \\
&  &  & Cloud &  &  & Cloud \\
\begin{tabular}[c]{@{}l@{}}Q1: It felt like i was in control of the body I was seeing (agency)\end{tabular} & 5(1) & 5(1) & 5(1) & 5(2) & 4(2) & 5(2) \\
\begin{tabular}[c]{@{}l@{}}Q2: It felt that the virtual body was my own body (Body Ownership)\end{tabular} & 4.5(2) & 4(1) & 5(3) & 4(3) & 4(2) & 5(2) \\
\begin{tabular}[c]{@{}l@{}}Q3: It felt as if my body was located where I saw the virtual body to be\end{tabular} & 5(2) & 5(3) & 4(2) & 3(3) & 2.5(2) & 3(3) \\
\begin{tabular}[c]{@{}l@{}}Q4: It felt as   if I had more than one body\end{tabular} & 2(2) & 2(2) & 2.5(3) & 3(2) & 3.5(2) & 4(2) \\
\begin{tabular}[c]{@{}l@{}}Q5: It was easy to walk on the virtual environment\end{tabular} & 5(2) & 5(1) & 5(1) & 5(2) & 4(2) & 5(2) \\
\begin{tabular}[c]{@{}l@{}}Q6: It was easy to avoid obstacles in the virtual environment (Task 1)\end{tabular} & 5(1) & 5(1) & 5(2) & 5(1) & 4(2) & 4.5(1) \\
\begin{tabular}[c]{@{}l@{}}Q7: It was easy to go over the obstacles in the virtual environment (Task 2)\end{tabular} & 5(1) & 5(2) & 5(1) & 5(1) & 4(2) & 5(1) \\
\begin{tabular}[c]{@{}l@{}}Q8: It was easy to go under the obstacles in the virtual environment (Task 3)\end{tabular} & 6(1) & 5(1) & 5(1) & 4(2) & 4(2) & 4(2) \\
\begin{tabular}[c]{@{}l@{}}Q9: It was easy to reach the thrown objects (Task 4)\end{tabular} & 4(2) & 4(2) & 4.5(2) & 4(1) & 4(3) & 2.5(2) \\
\begin{tabular}[c]{@{}l@{}}Q10: It was each to reach the red and green objects\end{tabular} & 6(1) & 6(1) & 6(1) & 5(1) & 5(2) & 5(1) \\
\begin{tabular}[c]{@{}l@{}}Q11: I felt fatigued while doing the test\end{tabular} & 1(1) & 1(1) & 1(1) & 1(2) & 1(3) & 2(3) \\ \bottomrule
\end{tabular}
\end{table*}

\section{Results and Discussion}

In this section, we present the main observations made during the tests as well as the difficulties and suggestions from users about the test task.
To assess the difference between the three user embodied representation both in first and third-person perspective, we collected both objective and subjective data, in the form of logs and inquiries respectively, during the evaluation sessions. 
For the continuous variable, i.e. time, we used Shapiro-Wilk test to assess data normality.
Since all samples were normally distributed, we used the Repeated Measures ANOVA with the Bonferroni correction applied to post-hoc tests for multiple comparisons, and the Paired-Samples T-Test test between two samples, to find statistically  significant differences.
For discrete data, such as the number of obstacle collisions or responses to the questionnaire, we resorted to the Friedman non-parametric test with Wilcoxon-Signed Ranks post-hoc test, also with a Bonferroni correction.

In the following subsections we present the analysis made based on the results of the questionnaires and log files data obtained during the test.

\subsection{User preferences}

We made two different comparisons based on the data collected through the questionnaires, between representations on the same perspective and representations between perspectives.

\subsubsection{Perspective} 
When comparing between representations in the First-Person Perspective, we found no statistical differences in the sense of embodiment, easiness to complete the task and fatigue felt during the test.
The only exception was on the Q8, where users felt that it was easier to perform the second task (go under the obstacle) with the Abstract representation in comparison with the Point-Cloud representation (Z=-2.64, p\textless0.05).

When comparing between representations in the Third-Person Perspective, the results were slightly better.
Users attributed a higher sense of embodiment to the Point-Cloud Avatar when comparing to the Mesh Avatar, with statistical significance only on the sense of agency (Z=-2.812,p=0\textless0.01), i.e. sense of control of the virtual body.
Even though, users felt more fatigue with the Point-Cloud avatar against the Mesh Avatar (Z=-2.53,p=0.011), on the Third-Person Perspective.

\subsubsection{Representation}

On the Mesh Avatars, users felt a stronger sense of embodiment in the First-Person Perspective in all its sub-components: sense of agency (Z=-2.687,p=0.007), body ownership(Z=-2.775,p=0.006) and self-location (Z=-3.574,p\textless0.005). They also found easier to walk in the virtual environment using 1PP (Z=-3.352,p=0.001).
About task easiness, they overall preferred the Mesh Avatar in First-Person but with statistical significance only in Tasks 1 (Z=-2.902, p=0.004) and 3(Z=-3.579,p\textless0.005), Q6 and Q8 respectively. 

Point-Cloud Avatars had a similar embodiment feeling in both perspectives, but users found easier to perform the third and fourth tasks, Q8 (Z=-2.771,p=0.006) and Q9 (Z=-3.695,p\textless0.005) respectively.

The Abstract representation had stronger sense of body-ownership and self-location, but with statistical significance only over self-location (Z=-3.422,p=0.001).
Users also found easier to walk around the Virtual Environment (Z=-2.838,p=0.005), avoid obstacles in the First (Z=-2.638,p=0.005) and Third (Z=-3.879,p\textless0.005) Tasks and on the reaching task (Z=-2.676,p=0.007).

\subsection{Task performance}

\begin{figure}
    \centering
    \includegraphics[width=\columnwidth,trim=2.75cm 13.5cm 4cm 2.5cm,clip=true]{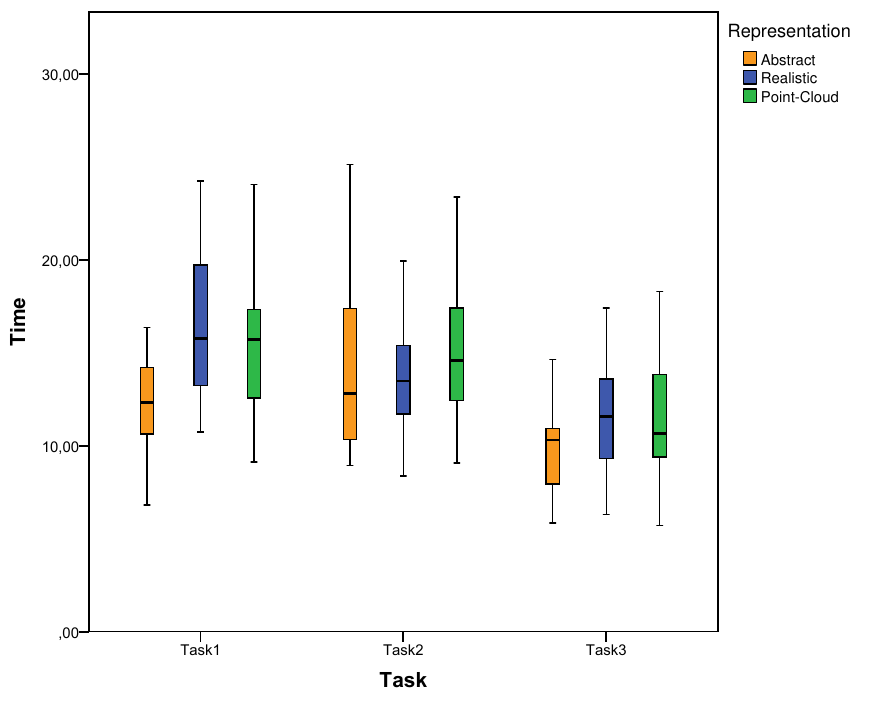}
    \caption{Performance time of Avatars in First-Person Perspective grouped by Task.  median, first and third interquartile ranges
(boxes) and 95\% confidence interval (whiskers). Orange represents the Abstract avatar, Blue the Realistic Mesh Avatar and Green, Point-Cloud Avatar.}
    \label{fig:1PPtimeByTask}
\end{figure}

\begin{figure}
    \centering
    \includegraphics[width=\columnwidth,trim=2.75cm 13.5cm 4cm 2.5cm,clip=true]{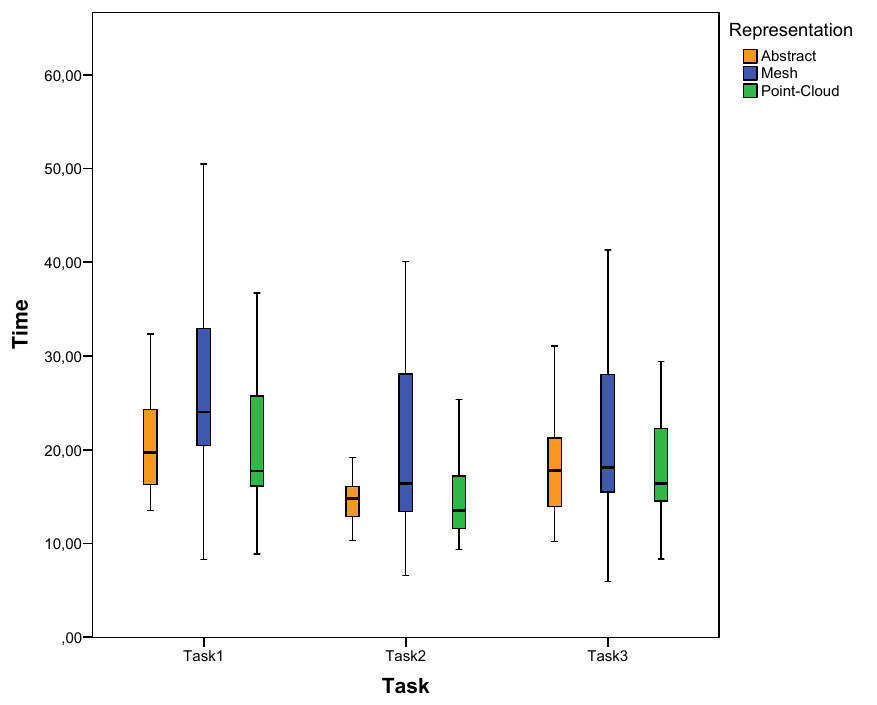}
    \caption{Performance time of Avatars in Third-Person Perspective grouped by Task.  median, first and third interquartile ranges
(boxes) and 95\% confidence interval (whiskers). Orange represents the Abstract avatar, Blue the Realistic Mesh Avatar and Green, Point-Cloud Avatar.}
    \label{fig:3PPtimeByTask}
\end{figure}

In this subsection we present the analysis of the results gathered from the users on the evaluation session. 
For assessing task performance of the users between the different representations we collected data through logs.
This data were: time, for evaluating efficiency of the representation; number of obstacles hit ,for spatial awareness evaluation.
Figures ~\ref{fig:1PPtimeByTask} and \ref{fig:3PPtimeByTask} show the times for each task in both perspectives and representations.
Since the fourth sub-task had a fixed time of execution we chose to not use time for this task in particular.
Instead, we used the number of balls caught during the fourth sub-task (Table~\ref{tab:hits}).
The number of obstacles hit and balls caught can be found on Table ~\ref{tab:collisions}. Because of the small amount of obstacle in each task we also used the percentage of collision avoidance as an additional metric in the comparison between representations. 

\begin{table*}[]
\centering
\caption{Task performance results. Median number of obstacles hit  inter-quartile range and percentage of obstacles avoided for tasks 1 to 3}

\begin{tabular}{@{}l|llllllllllll@{}}
\midrule
         & \multicolumn{6}{l}{First-Person Perspective(1PP)}             & \multicolumn{6}{l}{Third-Person Perspective(3PP)}            \\ \midrule
         & Abstract  &   & Mesh     &      & Point-Cloud  &         & Abstract  &  & Mesh &      & Point-Cloud &     \\ \midrule
Task \#1 & 2(2) & 33.33\% & 2(1) & 16.67\% & 2(2) & 4.17\%  & 1(2) & 33.33\% & 2(1) & 20.85\% & 0(2) & 54.17\% \\
Task \#2 & 2(1) & 20.83\% & 2(1) & 25.00\% & 3(2) & 16.67\% & 3(0) & 4.17\%  & 2(2) & 29.17\% & 3(0) & 0.00\%  \\
Task \#3 & 0(0) & 87.50\% & 0(0) & 87.50\% & 0(1) & 58.33\% & 1(1) & 50.00\% & 0(1) & 58.33\% & 0(1) & 79.17\% \\ \bottomrule
\end{tabular}
\label{tab:collisions}
\end{table*}

\begin{table}[]
\centering
\caption{Task Performance results. Median and inter-quartile range for the number of balls caught (Task 4)}
\label{tab:hits}
\begin{tabular}{@{}llll@{}}
\toprule
 & Abstract & Mesh & Point-Cloud \\ \midrule
First Person & 2.5(2) & 2.5(0) & 3(1) \\
Third Person & 2(1) & 2(1) & 0(1) \\ \bottomrule
\end{tabular}
\end{table}

In the following sub-sections we present the results obtained for each of the metrics used (time, number of obstacles hit and balls caught) for each of the sub-tasks.

\subsubsection{Task 1}

\textbf{Collision:} Grouping results by perspective we can say that users collided with more objects in the First-Person Perspective with the Point-Cloud avatar against the Abstract Avatar (Z=-2.668,p=0.008). In Third-Person Perspective we found a statistical significance between the Mesh and the Point-Cloud representation, with the Point-Cloud avatar having a higher number of obstacles avoided (Z=-2.542,p=0.011).

When grouping perspectives by representations we can say that we only found statistical significance in the Point-Cloud representation, with the Third-Person Perspective having the higher number of obstacles avoided (Z=-3.490,p\textless0.005).

\textbf{Time:} On the execution of the first task, we found a statistically significant difference in the First-Person Perspective between the Point-Cloud and the Abstract Avatar (p=0.024), with the abstract having the better performance. 
It was also found between the Abstract and Mesh Avatars (p=0.015), with the Abstract having the edge. 
When comparing the same representation for both perspectives, statistical significance was found between all avatars ( Abstract: t(18)=-6.312, p =0; Cloud: t(20)=-3.254, p= 0.001; and Mesh: t(16)=-3.254, p = .005). 
For the three representations, the First-Person Perspective had the better performance.

\subsubsection{Task 2}

\textbf{Collision:} In the representation-perspective test, we found statistical significance, with the Abstract avatar (Z=-2.714,p=0.007) having the most number of obstacles avoided overall in 3PP. 
Also, between the Mesh and Point-Cloud Avatars (Z=-3.779,p\textless0.005) and between Abstract and Mesh Avatars in the Third-Person Perspective. In the First-Person Perspective no statistical significant difference was found. 
No significance was found between 1PP representations.

Also, when comparing the different representations in both perspectives we found statistical significance for both Point-Cloud (Z=-2.941,p=0.003) and Mesh (Z=-3.673,p\textless0.005) Avatars, with the first person having the advantage in both cases.

\textbf{Time:} No statistically significant difference was found between any of the representation-perspective combinations.
When comparing the same representation and two different perspectives, only in the mesh representation a statistically significant difference was found (t(18)=-2.479, p=0.023), with the first person perspective having the advantage.

\subsubsection{Task 3}

\textbf{Collision:} In the task, no statistical significant difference was found when using the representation-perspective grouping factor. 
However, we found statistical significance in favor of the Abstract Avatar in 1PP against the same representation in 3PP (Z=-2.714,p=0.007).

\textbf{Time:} 
On the third task, we found a statistically significant difference when comparing the abstract (p=0.05) and the point cloud (p=0.045) representations to the mesh, but not between themselves.
In both situations, the mesh avatar had a worse performance.
Similarly to the first task, all representations had a better performance in the first person perspective (Abstract: t(20)=-6.76, p=\textless0.005; Mesh t(18)=-4.276; Point Cloud: t(19)=-5.354, p\textless0.005)

\subsubsection{Task 4}

For the ball catching task we found statistical significance on the First-Person the between Point-Cloud and Mesh Avatars (Z=-2.546,p=0.011),with better results for the Point-Cloud representation, and Abstract and Mesh Avatars (Z=-2.401)
Table \ref{tab:hits}, with better for the Mesh Avatar. 

When comparing the representations between perspectives, we found statistically better results for the Point-Cloud in Third-Person when compared with the same representation in the First-Person Perspective (Z=-3.961,p\textless0.005).

\subsection{Discussion}

According to the questionnaires, we can conclude that the perspective has more effect than representation on embodiment and task execution, specially on the Abstract and Mesh representations.
Also, we can state that the sense of embodiment with the Point-Cloud avatar is similar with the Abstract avatar, but significantly better when compared to the Mesh Avatar representation when the avatar is viewed through a Third-Person Perspective.

The lower sense of embodiment detected on the 3PP mesh representation, can be explained by the Uncanny Valley effect.
While the point cloud directly maps the users body, both the Mesh and the Abstract representations can be considered simplifications, since not all body movements are directly mapped by the Microsoft Kinect (eg. detailed hand and head movement).
This simplification is accepted on the Abstract representation, while its effect is perceived as an error in the Mesh avatar. 

Regarding time efficiency on the execution of the proposed tasks, we found that the 1PP had a clear advantage over the 3PP.
We argue that this result is related to the overall preference demonstrated by the users in the 1PP, as seen in Table \ref{tab:questionnaires}.
Being a more natural perspective to the user, the movement through the environment was faster when compared to the out of body experience of the 3PP.

When comparing different avatars, although we only found statistical significance in certain task-representation combinations, some considerations can be made. 
For the 1PP, the Abstract Avatar had the overall best performance (with statistical significance on the first task). 
Being a more minimalistic representation than the alternatives, less occlusion between body and obstacles were seen, with less distractions from the proposed goal. 

For the 3PP, the Mesh Representation had the overall worse performance (with statistical significance on the third task). 
We relate this result to the lower sense of embodiment seen in Table \ref{tab:questionnaires}and its relation to the Uncanny Valley effect, as explained above.
This was noticed to frequently slow down the interaction.

Regarding the successful completion of the proposed tasks, the only task where the 3PP had the overall advantage was the first task, where spatial awareness was the key factor.
Table \ref{tab:collisions} shows these results, with the Point-Cloud representation being the best performer between representations, with the lowest median number, and highest number of participants concluding the task without collisions (statistical significance found only when comparing with the mesh avatar).

For the second task, the 1PP had the advantage (statistical significance found for Mesh and Point-Cloud Avatar), except for the Point-Cloud representation, which had worse performance in both perspectives. 
We attribute this to the fact that this representation is visually richer. Although this helps with the sense of embodiment, occlusions are naturally created by the rendered splats, or clothing (pants, shoes) in both 3PP and 1PP and body parts (breasts, stomach) in 1PP.

The third task was highly influenced by the wrongful estimation of the height of the obstacle. 
This can explain the fact that no statistical significance was found between any representation/perspective, except for the Abstract Avatar performing better in 1PP.
Users would overestimate how low they needed to go to avoid the obstacle, having success most of the times.
However, when analysing the questionnaires, users reported an overall preference for the 1PP for this task.
This can be attributed to the fact that the virtual head in the 3PP is not exactly where the user's real head is.
The estimation of the height of the obstacle is more complicated from this point of view, and the user's sense of balance is more heavily affected by the use of the displaced camera in 3PP, since lowering your body can affect your sense of balance.

Finally, 1PP had the advantage on the fourth task (Table \ref{tab:hits}).
For this perspective, both the Point-Cloud and the Abstract Avatars had the advantage over the Mesh Avatar(with statistical significance).
For the 3PP, Point-Cloud avatars had the worst performance (statistical significance found).
We again attribute this to occlusions created by the users body and clothing which could inhibit the accurate visualization of the balls' trajectories.

We noticed that the most ``realistic" combination proposed (1PP Point-Cloud representation) had the fastest overall times, while having a high number of obstacles hit.  
This result can indicate that the user feels more confident in his distance estimation on this representation, leading him to a higher number of mistakes in avoiding obstacles.


\section{Conclusions}

The use of Head-Mounted Displays occludes the user's self, causing a decrease in the feeling of presence on the Virtual Reality session.
A way of overcoming this problem is by using self-embodied avatars which improve presence and overall distance estimation in VR setups.
Some factors that could influence the sense of embodiment are the realism and perspective which the avatar is viewed (in First or Third-Person Perspective).
The realism of avatars are affected by the Uncanny Valley which also influence self-embodied avatars.
Even though, the effects of both realism and perspective are not yet fully adressed in the literature when it comes to self-embodied avatars.
For that we used three different representations varying its realism in both perspectives (1PP and 3PP) following the Uncanny Valley effect, varying from an Abstract to a Realistic Representation. For the Realistic representation we used a Point-Cloud Avatar, which uses affordable depth-sensors to map real users' information into a user augmented-self inside the Virtual Environment.
To assess each of the representation-perspective combinations, we chose natural tasks such as walking while avoiding obstacles and catching thrown objects.

As a result of the statistical analysis, discussion and evaluation of the results, we propose the following guidelines regarding body representation and camera perspective for embodied virtual reality applications:

\begin{itemize}
    \item The uncanny valley effect is more prevalent on the third person perspective, and it influences the time efficiency of navigation tasks. 
    \item On a First-Person Perspective (1PP), the Uncanny Valley Effect is only noticed on tasks where a higher sense of embodiment is required (eg. reflex-based tasks).
    \item To avoid the Uncanny Valley Effect, one should use either a simplified (Abstract) or a Realistic (Point-Cloud) representation of the participant.
    \item Using a more realistic representation (Point-Cloud) and Perspective (1PP) can lead to faster execution times, but lower efficacy on the execution of tasks.
    \item A Third-Person Perspective is recommended when spatial awareness on the horizontal navigation plane is required, with a slight advantage when using a Realistic Representation.
    \item Obstacle avoidance and reaching moving objects can be problematic when using a Realistic representation due to occlusions created by the visually richer representation.
    \item The users' sense of balance is negatively affected by a Third-Person Perspective. 
    \item Reflex-based tasks have better performance when using a First-Person Perspective. 
\end{itemize}

However, we were not able to clarify certain aspects investigated in this study.
Namely, avoiding obstacles in the vertical plane, and how the poor estimation of distance can influence these tasks, and if the body representation has any effect on this matter.
Also, the effect of the Uncanny Valley on reflex-based tasks was noticed, but requires further evaluation using different stimuli.

Going forward, different tasks should be considered for the same perspectives and avatars, such as collaboration between different users, social environments, and communicative tasks.


\bibliographystyle{abbrv}
\bibliography{main}
\end{document}